# Study of the projectile motion in midair using simple analytical formulas

Peter Chudinov

*Department of Engineering, Perm State Agro-Technological University, 614990, Perm, Russia*

A classic problem of the motion of a projectile thrown at an angle to the horizon is studied. Air resistance force is taken into account with the use of the quadratic resistance law. The projectile motion is described analytically with fairly simple formulas. They make it possible to calculate basic motion characteristics and trajectory as easily as in the case of a parabolic motion. There is no need to study the problem numerically. The proposed formulas are universal, that is, they can be used for any initial conditions of throwing. In addition, they have acceptable accuracy over a wide range of the change of parameters. The motion of a baseball, a tennis ball and shuttlecock of badminton are presented as examples.

## 1. Introduction

The study of the motion of a projectile, thrown at an angle to the horizon, is a wonderful classical problem. This issue has been the subject of great interest for investigators for centuries. Currently, the study of parabolic motion, in the absence of any drag force, is a common example in introductory physics courses. The theory of parabolic motion allows you to analytically determine the trajectory and all important characteristics of the movement of the projectile. Introduction of air resistance forces into the study of the motion, however, complicates the problem and makes it difficult to obtain analytical solutions, except in a few particular cases. This especially applies to the movement of the projectile, subjected to quadratic air drag force. Numerical studies of projectile motion with quadratic dependence on projectile speed have been studied in many works. From an educational point of view, such studies suggest that students are confident in numerical methods. So the description of the projectile motion by means of simple approximate analytical formulas under air resistance has great methodological and educational importance.

Recently, studies of the movement of a projectile with a quadratic law of resistance have appeared, in which high-precision analytical formulas for describing such movement have been proposed. Using the proposed formulas, it is possible to describe the movement of a projectile with a quadratic resistance as completely and relatively simply as in the case of a parabolic motion. The main goal of this work is to give analytical formulas for the projectile trajectory and movement characteristics as simple as possible from a technical point of view, in order to be grasped even by first-year undergraduates. The conditions of applicability of the quadratic resistance law are deemed to be fulfilled, i.e. Reynolds number $Re$ lies within $1 \times 10^3 < Re < 2 \times 10^5$. Magnus forces are not included in this work.

## 2. Equations of projectile motion

Here we state the formulation of the problem and the equations of the motion [1], [4]. Let us consider the motion of a projectile with mass $m$ launched at an angle $\theta_0$ with an initial speed $V_0$ under the influence the force of gravity and resistance force $R = mgkV^2$. Here $g$ is the acceleration of gravity, $k$ is the drag constant and $V$ is the speed of the object. Air resistance force $R$ is proportional to the square of the speed of the projectile and is directed opposite the velocity vector. It is assumed that the projectile is at the origin at the initial instant and the point of impact of the projectile lies on the same horizontal $y = 0$

(see Fig. 1). In ballistics, the movement of a projectile is often studied in projections on natural axes [1]. The equations of the projectile motion in this case have the form

$$\frac{dV}{dt} = -g\sin\theta - gkV^2, \quad \frac{d\theta}{dt} = -\frac{g\cos\theta}{V}, \quad \frac{dx}{dt} = V\cos\theta, \quad \frac{dy}{dt} = V\sin\theta. \quad (1)$$

Here $\theta$ is the angle between the tangent to the trajectory of the projectile and the horizontal, $x, y$ are the Cartesian coordinates of the projectile, $k = 1/V_{term}^2 = const$, $V_{term}$ is the terminal velocity.

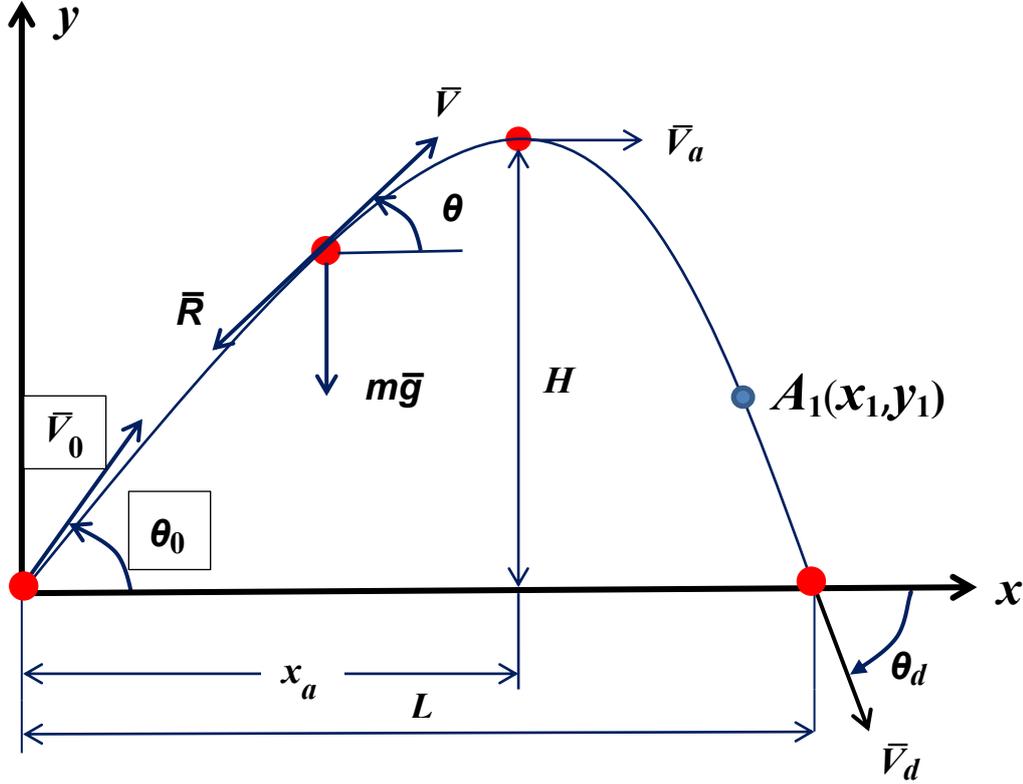

**Fig.1.** Basic motion parameters.

The well-known solution of system (1) consists of an explicit analytical dependence of the velocity on the slope angle of the trajectory and three quadratures

$$V(\theta) = \frac{V_0 \cos\theta_0}{\cos\theta\sqrt{1 + kV_0^2 \cos^2\theta_0 \left(f(\theta_0) - f(\theta)\right)}}, \quad f(\theta) = \frac{\sin\theta}{\cos^2\theta} + \ln\tan\left(\frac{\theta}{2} + \frac{\pi}{4}\right), \quad (2)$$

$$x = x_0 - \frac{1}{g}\int_{\theta_0}^{\theta} V^2 d\theta, \quad y = y_0 - \frac{1}{g}\int_{\theta_0}^{\theta} V^2 \tan\theta d\theta, \quad t = t_0 - \frac{1}{g}\int_{\theta_0}^{\theta} \frac{V}{\cos\theta} d\theta. \quad (3)$$

Here $t_0$ is the initial value of the time, $x_0, y_0$ are the initial values of the coordinates of the projectile (usually accepted $t_0 = x_0 = y_0 = 0$). The derivation of the formulae (2) is shown in the well-known monograph [2]. The integrals on the right-hand sides of formulas (3) cannot be expressed in terms of elementary functions. Hence, to determine the variables $t, x$ and $y$ we must either integrate system (1) numerically or evaluate the definite integrals (3).

### 3. Calculation formulas of the problem

When the projectile moves, we are most interested in the basic geometric and temporal characteristics of the motion and its trajectory. This article provides formulas for determining the following seven

characteristics of the movement:

    $H$ - maximum height of ascent of the projectile,
    $T$ - motion time,
    $L$ - flight range,
    $x_a$ - the abscissa of the trajectory apex,
    $t_a$ - the time of ascent,
    $x_{as}$ - asymptote of the projectile trajectory,
    $\theta_d$ - impact angle with respect to the horizontal (see Fig. 1).

The trajectory equation in this paper is written in the same way as in the parabolic theory, that is, in the form of an explicit function $y(x)$.

This article is based on two recent studies [3], [5] of the movement of a projectile with a square law of resistance of the medium and on an earlier paper [4]. In [3], a very fruitful idea was proposed for calculating integrals (3) and high-precision formulas were found for calculating the parameters $H, x_a, t_a$. Based on this idea, in [5] high-precision analytical solutions were received for the basic functional dependencies of the problem $x(\theta), y(\theta), t(\theta)$ in terms of elementary functions. These functions allow us to determine the values of coordinates $x, y$ and time $t$ at any point on the trajectory, that is, for any value of the inclination angle $\theta$. Based on the results of [3], [5], we write formulas for the characteristics $H, x_a$ and $t_a$:

$$x_a = \frac{2}{gk\alpha_1 b_3}\arctan\left(\frac{b_3}{2b_1\cot\theta_0 - 1}\right), \quad H = -\frac{\ln\lambda_1}{gk\alpha_2} - \frac{x_a}{2b_2},$$

$$t_a = \frac{1}{g\sqrt{k\alpha_2}}\arccos\left(\frac{1 + 2b_2\tan\theta_0 + 4b_1 b_2 \lambda_1}{1 + 4b_1 b_2}\right). \tag{4}$$

The following notations are introduced here:

$$\alpha_1 = 2\cot\theta_0 \ln\tan\left(\frac{\theta_0}{2} + \frac{\pi}{4}\right), \quad \alpha_2 = \frac{1}{\sin\theta_0} - \frac{\alpha_1 \cot\theta_0}{2}, \quad b_1 = \frac{1 + kV_0^2\sin\theta_0}{\alpha_1 kV_0^2 \cos^2\theta_0} + \frac{\tan\theta_0}{2},$$

$$b_2 = \frac{\alpha_2}{\alpha_1}, \quad b_3 = \sqrt{-1 - 4b_1 b_2}, \quad b_4 = \sqrt{-1 + 4b_1 b_2}, \quad \lambda_1 = \sqrt{1 - \frac{\tan\theta_0}{b_1}(1 + b_2 \tan\theta_0)}.$$

In [4] an equation was proposed for the projectile trajectory in the form of explicit function $y(x)$ of the following kind

$$y(x) = \frac{Hx(L - x)}{x_a^2 + (L - 2x_a)x}. \tag{5}$$

From the structure of equation (5) it follows that for its application it is necessary to know three parameters – $H, x_a, L$. The parameters $H$ and $x_a$ are determined by formulas (4). To determine $L$, we use the same trajectory equation (5). For this purpose, we take some point $A_1$ with coordinates $x_1, y_1$ on the trajectory. Coordinates $x_1, y_1$ are given by the relations $x_1 = x(-\theta_1)$, $y_1 = y(-\theta_1)$, where $\theta_1 = \theta_0/2 + \pi/4$. The value $\theta_1$ of the angle of inclination ensures the location of point $A_1$ near the point of impact of the projectile (see Fig. 1). For small angles of throw, point $A_1$ lies above the point of impact, for large angles of throw point $A_1$ lies below the point of impact. We recall that in [5] the equation of the trajectory was obtained in the parametric form $x(\theta), y(\theta)$. These functions allow us to compute values $x_1, y_1$. We substitute the coordinates $x_1, y_1$ in relation (5) and will consider this relation as an equation for the unknown quantity $L$:

$$y_1 = \frac{Hx_1(L-x_1)}{x_a^2+(L-2x_a)x_1}.$$

From here we get

$$L = x_1\left(1+\frac{(\lambda_2-1)^2}{\lambda_3-1}\right), \quad \text{where } \lambda_2 = \frac{x_a}{x_1}, \quad \lambda_3 = \frac{H}{y_1}. \qquad (6)$$

Here $x_1, y_1$ are the coordinates of the point $A_1$:

$$x_1 = x_a + \frac{2}{gk\alpha_1 b_4}\arctan\left(\frac{b_4}{1+2b_1\cot\theta_1}\right), \quad y_1 = H + \frac{x_1-x_a}{2b_2} - \frac{1}{2gk\alpha_2}\ln\left|\frac{b_1+\tan\theta_1+b_2\tan^2\theta_1}{b_1}\right|.$$

Having determined the parameter $L$, we can use function (5) to construct the trajectory of the projectile. When moving in a medium with resistance, the trajectory of the projectile has a vertical asymptote. The formula for calculating the value of the vertical asymptote $x_{as}$ of the projectile trajectory is taken from [5]:

$$x_{as} = x_1 + \frac{2}{gk\beta_1 d_2}\text{arccot}\left(\frac{1+2d_1\tan\theta_1}{d_2}\right). \qquad (7)$$

The following notations are introduced here:

$$d_0 = \frac{b_1\alpha_1-\beta_0}{\beta_1}, \quad d_1 = \frac{\beta_2}{\beta_1}, \quad d_2 = \sqrt{4d_0 d_1 - 1}, \quad \beta_2 = \frac{f(89°)-f(\theta_1)}{(\tan 89°-\tan\theta_1)^2} - \frac{2}{\cos\theta_1(\tan 89°-\tan\theta_1)},$$

$$\beta_1 = \frac{2(1-\beta_2\sin\theta_1)}{\cos\theta_1}, \quad \beta_0 = \beta_1\tan\theta_1+\beta_2\tan^2\theta_1-f(\theta_1).$$

The function $f(\theta)$ is defined by the relation (2). The projectile travel time $T$ and the angle of fall are defined by formulas from [4]

$$T = 2\sqrt{\frac{2H}{g}}, \quad \theta_d = -\arctan\left(\frac{HL}{(L-x_a)^2}\right). \qquad (8)$$

By analogy with the function $y(x)$ of the form (5), we can construct an explicit function $y(t)$ of the same form

$$y(t) = \frac{Ht(T-t)}{t_a^2+(T-2t_a)t}. \qquad (9)$$

Thus, a relatively simple set of formulas (4) – (8) has been obtained that makes it possible to study the motion of a projectile in a medium with a quadratic law of resistance. Motion characteristics $L, x_a, t_a, H, T, \theta_d, x_{as}$ are calculated by formulas (4), (6), (7), (8) and formula (5) describes the trajectory of the projectile. These formulas are universal. They are applicable for any initial conditions of projectile throwing and for any values of the drag coefficient $k$ and do not require knowledge of any numerical methods.

## 4. Results of the calculations

Let us apply the obtained formulas to study the motion of sports equipment – a baseball, a tennis ball, and a badminton shuttlecock. In all calculations, the value $g$ is $g = 9.81$ m/s$^2$. The drag coefficient $k$, used in formulas (4) – (8), can be calculated through the terminal speed: $k = 1/V_{term}^2$. According to the data of [6], we have for baseball $V_{term} = 40$ m/s, for tennis ball $V_{term} = 22$ m/s, for badminton shuttlecock

$V_{term} = 6.7$ m/s. Then, respectively, we have, for baseball $k = 0.000625$ s$^2$/m$^2$, for tennis ball $k = 0.002$ s$^2$/m$^2$, for shuttlecock $k = 0.022$ s$^2$/m$^2$. Note that the coefficient $k$ for baseball and shuttlecock is 35 times different. The resistance coefficient $k$ can also be determined as follows

$$k = \frac{\rho_a c_d S}{2mg} = const,$$

$\rho_a$ is the air density, $c_d$ is the drag factor for a sphere, $S$ is the cross-section area of the object. In the calculations, we use the typical speeds of these projectiles – for baseball $V_0 = 55$ m/s, for tennis ball $V_0 = 70$ m/s, for shuttlecock $V_0 = 80$ m/s. Throw angle values for curves numbered 1, 2, 3 on the graphs, are respectively $\theta_0 = 20°, 45°, 70°$.

Fig. 2,3 presents the results of plotting the projectile trajectories with the aid of formulas (4) – (8) over a wide range of the change of the initial angle $\theta_0$.

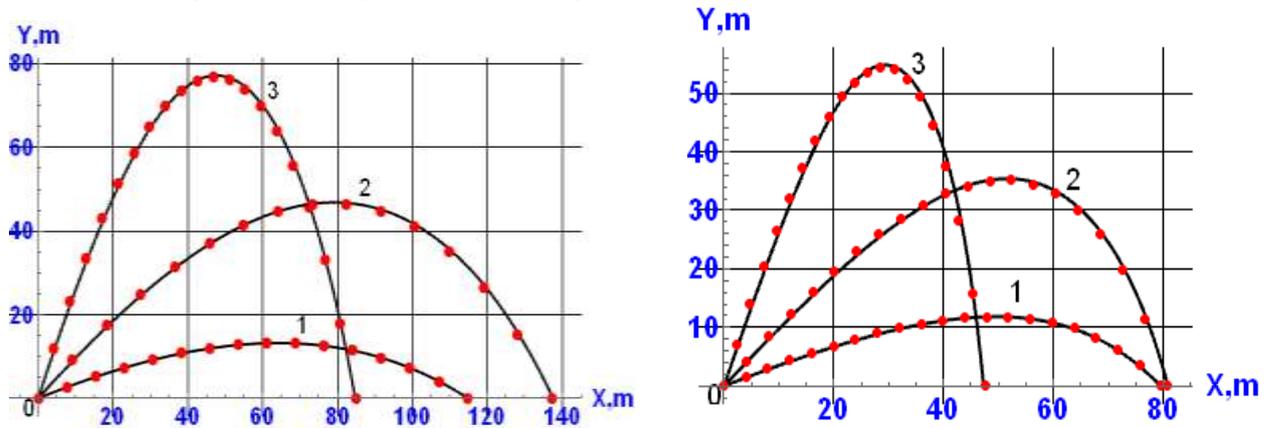

**Fig. 2.** The graphs of the trajectory $y = y(x)$ for baseball (left) and tennis (right).

The thick solid black lines in Fig. 2,3 are obtained by numerical integration of system (1) with the aid of the 4-th order Runge-Kutta method. The red dot lines are obtained using analytical formulas (4) – (8). As it can be seen from Fig. 2,3, formulas (4) – (8) with high accuracy approximate the trajectory of the projectile in a fairly wide range of angle $\theta_0$ and in a wide range of variation of the drag coefficient $k$.

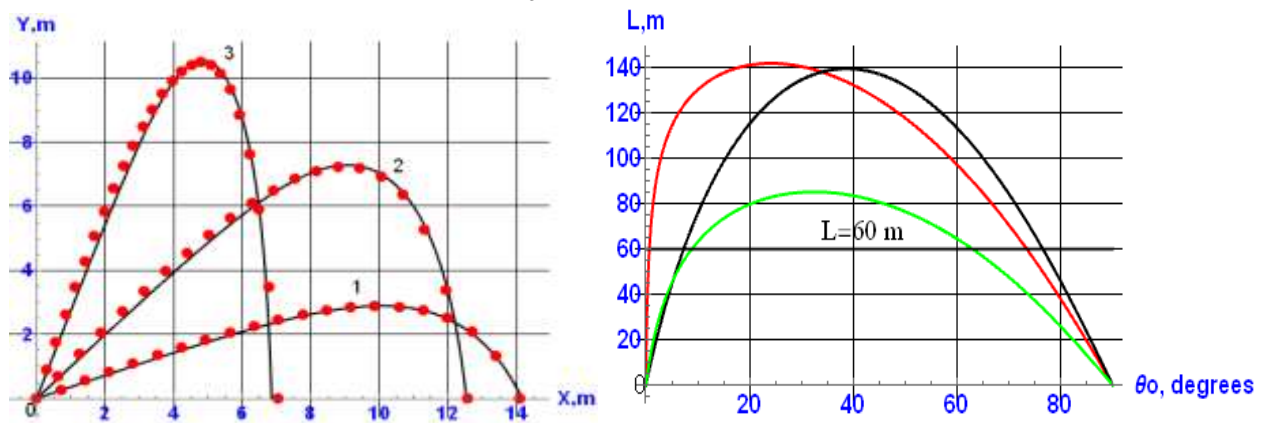

**Fig. 3.** The trajectories $y(x)$ for shuttlecock.  **Fig. 4.** The graphs of the function $L(\theta_0)$.

We pay a special attention to Fig. 3. Trajectory number 2 is plotted for the initial conditions of throwing $V_0 = 80$ m/s, $\theta_0 = 45°$, $k = 0.022$ s$^2$/m$^2$. Parameters $H$ and $L$ calculated by formulas (4) and (6) are determined precisely: $H = 7.2$ m, $L = 12.5$ m. In the absence of air resistance ($k = 0$), according to the parabolic theory formulas these parameters would be as follows:

$$H = \frac{V_0^2 \sin^2 \theta_0}{2g} = 163.1 \text{ m}, \quad L = \frac{V_0^2 \sin 2\theta_0}{g} = 652.4 \text{ m}, \quad T = \frac{2V_0 \sin \theta_0}{g} = 11.533 \text{ s}. \quad (10)$$

Comparison of the values $H$ and $L$ when moving with resistance and without resistance shows the huge influence of air resistance on the motion of the shuttlecock.

Formulas (4), (6), (8) are completely different from parabolic theory formulas (10). Nevertheless, for very small values of the drag coefficient $k$, formulas (4), (6), (8) give absolutely the same values for parameters $H, L, T$ like the formulas of the parabolic theory. For example, for $V_0 = 80$ m/s, $\theta_0 = 45°$, $k = 10^{-13}$ s$^2$/m$^2$ formulas (4), (6), (8) give the values $H = 163.1$ m, $L = 652.4$ m, $T = 11.533$ s. The value $k = 0$ cannot be used in formulas (4), (6), (8), since division by zero occurs.

Table 1 contains the values of the motion characteristics $L, x_a, t_a, H, T, \theta_d, x_{as}$, calculated by formulas (4) – (8) and corresponding to Fig. 2 and 3. The analytical values are compared with the numerical values found by integration of system (1). The top row of each cell in Table 1 contains the values of the corresponding parameter for baseball, the middle row contains the parameter values for shuttlecock, the bottom row contains the parameter values for tennis. The table data indicate good accuracy of the formulas (4) – (8).

**Table 1. Basic motion parameters of the baseball, tennis and shuttlecock.**

| | $\theta_0 = 20°$ | | | $\theta_0 = 45°$ | | | $\theta_0 = 70°$ | | |
|---|---|---|---|---|---|---|---|---|---|
| | num. value | analytical value | error (%) | num. value | analytical value | error (%) | num. value | analytical value | error (%) |
| $L$, m | 114.88 | 115.17 | 0.3 | 137.63 | 137.04 | -0.4 | 84.72 | 84.22 | -0.6 |
| | 14.17 | 14.10 | -0.5 | 12.54 | 12.57 | 0.2 | 6.86 | 7.05 | 2.8 |
| | 79.46 | 79.69 | 0.3 | 80.86 | 80.62 | -0.3 | 47.27 | 47.58 | 0.7 |
| $x_a$, m | 64.13 | 64.08 | -0.1 | 79.05 | 78.57 | -0.6 | 47.6 | 47.2 | -0.8 |
| | 10.06 | 10.05 | -0.1 | 9.07 | 9.01 | -0.7 | 4.84 | 4.80 | -0.8 |
| | 49.48 | 49.40 | -0.2 | 51.62 | 51.18 | -0.9 | 29.32 | 29.04 | -1.0 |
| $H$, m | 13.22 | 13.21 | -0.1 | 46.84 | 46.61 | -0.5 | 77.20 | 76.74 | -0.6 |
| | 2.89 | 2.89 | 0 | 7.28 | 7.24 | -0.5 | 10.54 | 10.49 | -0.5 |
| | 11.82 | 11.81 | -0.1 | 35.39 | 35.15 | -0.7 | 54.81 | 54.42 | -0.7 |
| $T$, s | 3.263 | 3.282 | 0.6 | 6.137 | 6.166 | 0.5 | 7.932 | 7.911 | -0.3 |
| | 1.428 | 1.534 | 7.4 | 2.386 | 2.43 | 1.8 | 3.022 | 2.925 | -3.2 |
| | 3.020 | 3.103 | 2.7 | 5.275 | 5.355 | 1.5 | 6.712 | 6.662 | -0.7 |
| $\theta_d$, ° | -30.3 | -30.2 | -0.3 | -62.2 | -61.8 | -0.6 | -78.8 | -78.0 | -1.0 |
| | -65.5 | -68.0 | 3.8 | -84.0 | -82.1 | -2.3 | -88.3 | -86.1 | -2.5 |
| | -45.0 | -45.7 | 1.5 | -74.1 | -73.0 | -1.5 | -84.2 | -82.4 | -2.1 |
| $t_a$, s | 1.532 | 1.532 | 0 | 2.801 | 2.792 | -0.3 | 3.603 | 3.587 | -0.4 |
| | 0.518 | 0.517 | -0.1 | 0.793 | 0.789 | -0.5 | 0.967 | 0.961 | -0.6 |
| | 1.303 | 1.301 | -0.2 | 2.164 | 2.153 | -0.5 | 2.707 | 2.690 | -0.6 |
| $x_{as}$, m | 234.0 | 233.8 | -0.1 | 199.0 | 198.8 | -0.1 | 111.6 | 111.5 | -0.1 |
| | 15.9 | 15.9 | 0 | 13.0 | 13.0 | 0 | 7.0 | 7.0 | 0 |
| | 111.4 | 111.3 | -0.1 | 93.7 | 93.5 | -0.2 | 52.2 | 52.0 | -0.4 |

One of the important tasks of studying motion is to find the optimal throw angle $\theta_0^{opt}$, that maximizes the range. The proposed formulas (4) – (8) make it easy to solve this problem. For this purpose we note, that the range $L$ determined by formula (6) is found as a function of the form $L = L(\theta_0, V_0, k)$. It is enough to plot the graph of this function on the interval $0° \leq \theta_0 \leq 90°$ at a fixed value $V_0$. From this graph, both the optimal throw angle $\theta_0^{opt}$ and the maximum range $L_{max}$ are determined. In addition, the straight line $L = const$ on this graph allows us to determine two initial throwing angles which give the same range. Fig. 4 contains the graphs of the function $L(\theta_0)$. The black line is for baseball, the green line is for tennis, the red line is for shuttlecock. The $L$ value for shuttlecock is increased 10 times. The values of the initial velocities $V_0$ are the same as in Fig. 2,3. Comparison of the results of Fig. 4 with the numerical results obtained by integrating equations (1) with the aid of the 4-th order Runge-Kutta method is given in Table 2. Straight line $L = 60$ m in Fig. 4 allows us to determine the corresponding throw angles of $\theta_0 = 9°$ and $\theta_0 = 63°$ for tennis, $\theta_0 = 7°$ and $\theta_0 = 77°$ for baseball.

**Table 2. Optimal launching angle and maximal range**

|  | baseball | | | tennis | | | shuttlecock | | |
|---|---|---|---|---|---|---|---|---|---|
|  | num. value | analytical value | error (%) | num. value | analytical value | error (%) | num. value | analytical value | error (%) |
| $\theta_0^{opt}$, ° | 38.9° | 38.7° | -0.5 | 32.6° | 32.4° | -0.6 | 24.1° | 24.1° | 0 |
| $L_{max}$, m | 139.8 | 139.4 | -0.3 | 85.4 | 85.1 | -0.4 | 14.3 | 14.2 | -0.7 |

Note the remarkable accuracy of determining the optimal throwing angle of the projectile and maximum range. With an increase of the drag coefficient $k$, the optimal throwing angle decreases (especially significant for shuttlecock).

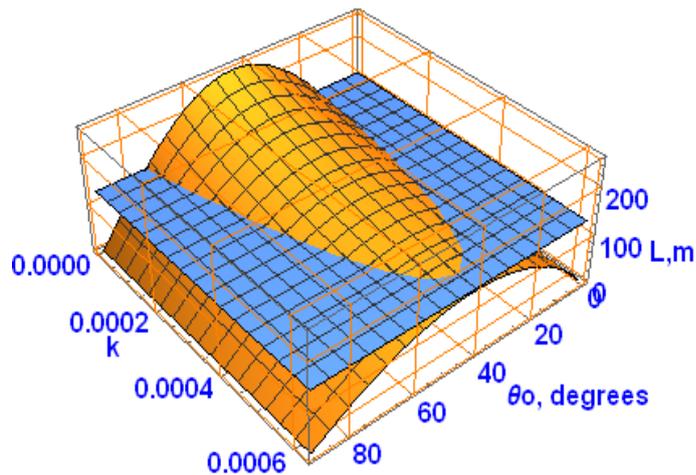
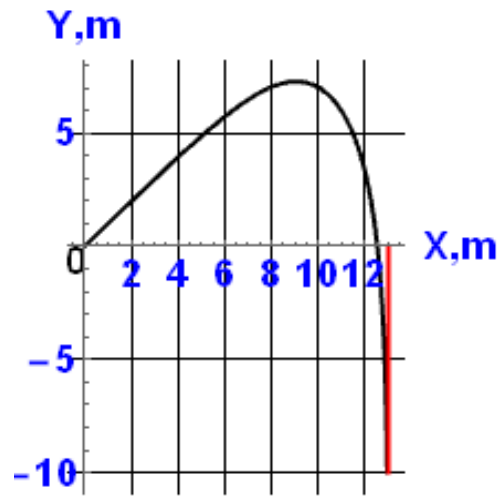

**Fig. 5.** Surface $L = L(\theta_0, k)$.   **Fig. 6.** The trajectory $y(x)$ for shuttlecock with asymptote.

In Fig. 5, the surface $L = L(\theta_0, k)$ is constructed for values $g = 9.81$ m/s$^2$, $V_0 = 50$ m/s. The intersection of the plane (blue) $L = 130$ m and the surface $L = L(\theta_0, k)$ in Fig. 5 allows us to determine the values of the parameters $\theta_0, k$ that ensure the achievement of the range $L = 130$ m. The trajectory $y(x)$ of the shuttlecock with asymptote is shown in Fig. 6 for $\theta_0 = 45°$. The thick solid black line in Fig. 6 is obtained by numerical integration of system (1) with the aid of the 4-th order Runge-Kutta method. The $x_{as}$ value was calculated using the formula (7). The asymptote is shown by a solid vertical red line.

Fig. 7 is a kind of nomogram. In this figure on the parameters plane ($L_{max}$, $\theta_0^{opt}$) seven lines for various values of the initial velocity $V_0$ and the drag coefficient $k$ are plotted. All points of these lines correspond to certain values of the parameters $V_0$, $k$. Line number 0 is drawn at $k_0 = 0.000625$ s$^2$/m$^2$, line 1 at $1.5k_0$, line 2 – at $2k_0$, line 3 – at $3k_0$, line 4 – at $4k_0$, line 5 – at a value of $6k_0$, line 6 – at a value of $10k_0$. Each line contains 8 points, each of which is plotted for a certain value of the initial velocity $V_0$. The first point (counting from the lower right corner of Fig. 7) is plotted at the value $V_0 = 10$ m / s, the second point - at the value $V_0 = 20$ m / s, etc. The values of $L_{max}$ and $\theta_0^{opt}$ for each point were calculated by formulas (4) − (6) by plotting the graphs of the function $L = L(\theta_0, V_0, k)$. This figure can be used for approximate finding the maximum range $L_{max}$ for the given values of the parameters $V_0$, $k$. Let it be required, for example, to determine the maximum range of the projectile at values

$$V_0 = 55 \text{ m/s}, \quad k = 0.0015 \text{ s}^2/\text{m}^2. \tag{11}$$

The value of the parameter $k$ lies between lines 2 and 3, the value of the parameter $V_0$ lies between points 5 and 6. Therefore, the values (11) in Fig. 7 approximately corresponds to point A. Dropping the perpendicular from point A to the axis $L_{max}$, we obtain an approximate value of this quantity for values (11): $L_{max}$ = 87 m. Dropping the perpendicular from point A to the axis $\theta_0^{opt}$, we obtain an approximate value of this parameter for values (10): $\theta_0^{opt}$ = 35.9 °. The exact values of the parameters $L_{max}, \theta_0^{opt}$, determined numerically using equations (1) for values (10), are as follows: $L_{max}$ = 86.8 m, $\theta_0^{opt}$ = 35.7°.

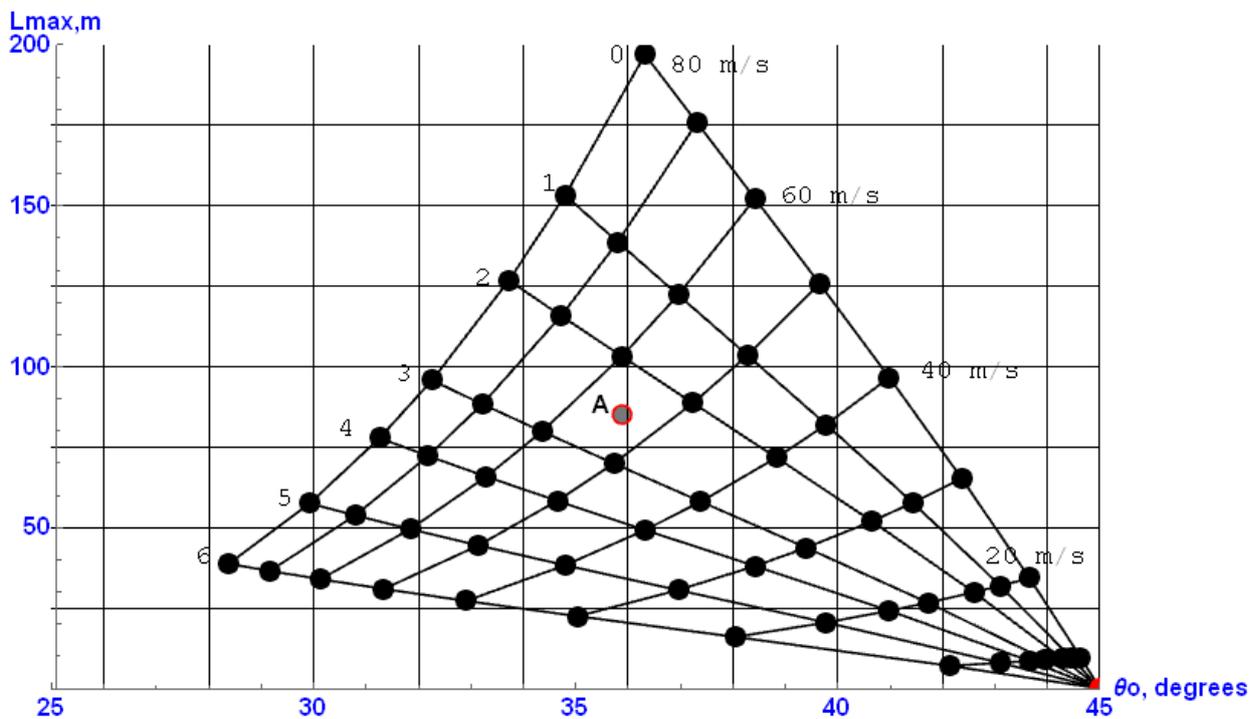

**Fig. 7.** Parameters plane ($L_{max}$, $\theta_0^{opt}$).

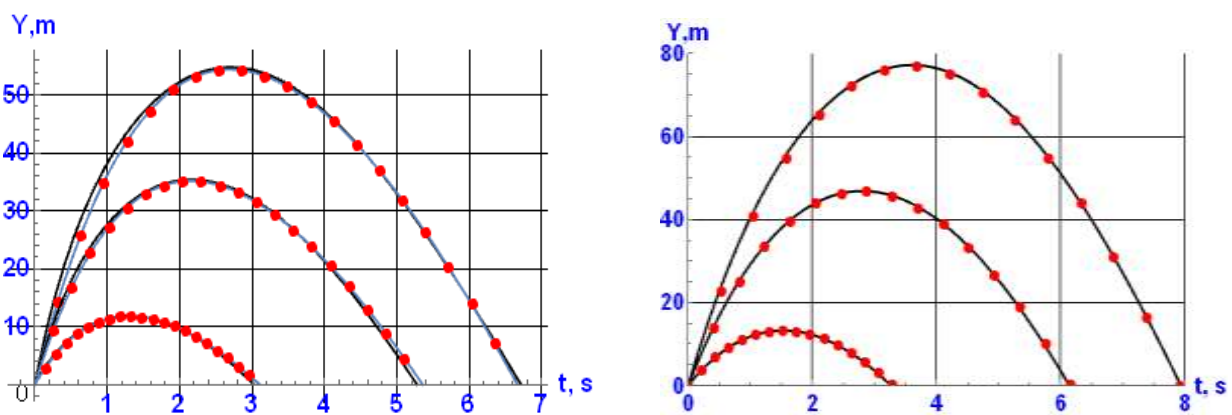

**Fig. 8.** The graphs of the function $y(t)$ for the tennis (left) and baseball (right).

The graphs of the function $y(t)$ constructed according to formula (9) are shown in Fig. 8 for a tennis ball and baseball. The used values $V_0$, $\theta_0$ are the same as in Fig. 2. The thick solid black lines in Fig. 8 are obtained with numerical integration of system (1) with the aid of the 4-th order Runge-Kutta method. The red dot lines are obtained using analytical formula (9). Fig. 8 shows that function (9) well approximates the dependence $y(t)$.

## 5. Conclusion

The proposed approach based on the use of relatively simple analytic formulae makes it possible to simplify significantly a qualitative analysis of the motion of a projectile with the air drag taken into account. All basic characteristics of the motion are described by non-complicated (simple) formulae containing elementary functions, as in the parabolic theory. To do this, it is enough to take the initial conditions of throwing $V_0, \theta_0$, the drag coefficient $k$ and substitute them in formulas (4) – (8). Formulas (4), (6), (7), (8) will determine the characteristics of the motion $L, x_a, t_a, H, T, \theta_d, x_{as}$, and formula (5) will determine the trajectory of projectile in the usual form $y = y(x)$. Moreover, the numerical values of these parameters are determined with an acceptable accuracy. The proposed analytical formulas can be useful for all researchers of this classical problem, even for people with high school math skills. We also note the universality of the proposed formulae, which do not have restrictions on the initial conditions and parameters.

## References

1. B. Okunev, *Ballistics, Vol. 2*, ( Voyenizdat, Moscow, 1943), p. 14
2. S. Timoshenko and D. Young *Advanced Dynamics,* (McGrow-Hill Book Company, New-York 1948), p. 112
3. M. Turkyilmazoglu, " Highly accurate analytic formulae for projectile motion subjected to quadratic drag," *Eur. J. Phys.* **37** 035001 (2016), https://doi.org/10.1088/0143-0807/37/3/035001
4. P. Chudinov, " Analytical investigation of point mass motion in midair," *Eur. J. Phys.* **25** 73-79 (2004), https://doi.org/10.1088/0143-0807/25/1/010
5. P. Chudinov, V. Eltyshev, Y. Barykin, "Analytical construction of the projectile motion trajectory in midair," *MOMENTO Rev. de Fis*. **62** 79-96 (2021), https://doi.org/10.15446/mo.n62.90752
6. C. Cohen, B. Darbois-Texier, G. Dupeux , E. Brunel, D. Quéré and C. Clanet, "The aerodynamic wall," *Proc. R. Soc. A* **470** 20130497 (2014), https://doi.org/10.1098/rspa.2013.0497